# Sorting and Routing Topologically Protected Edge States by Their Valleys

Kueifu Lai[1,3], Yang Yu[1], Yuchen Han[2], Fei Gao[4], Baile Zhang[4], and Gennady Shvets[1*]

[1]School of Applied and Engineering Physics, Cornell University, Ithaca NY 14853, USA.

[2]Department of Physics, Cornell University, Ithaca NY 14853, USA.

[3]Department of Physics, University of Texas at Austin, Austin, TX 78712, USA.

[4]Division of Physics and Applied Physics, School of Physical and Mathematical Sciences, Nanyang Technological University, Singapore 637371, Singapore.

* Author to whom correspondence should be addressed; Gennady Shvets: gs656@cornell.edu

**Abstract**

We experimentally demonstrate and characterize a photonic platform containing heterogeneous topological phases. This platform is shown to support valley-spin locked edge states that can be routed based on their spin and valley degrees of freedom. The routing is accomplished by changing the topology of the valley photonic crystal embedded inside a spin-Hall photonic crystal.

Topological insulators (TIs) [ 1,2,3,4,5,6,7] is one of the most popular fields in condensed matter physics for the past few decades. The photonic equivalent, namely photonic topological insulators (PTIs) [ 8,9,10,11,12,13,14,15,16,17], is also a fast-growing topic in the field of photonics in recent years due to the capability to emulate the electronic counterpart with ease. Given the fermionic nature of electrons is very different from bosonic photons, researchers used various constructs to grant the necessary properties in the realm of electromagnetism to realize the PTIs. One commonly used approach is to emulate the quantum Hall (QH) effect [ 18,19,20,21] by applying either external or effective magnetic field , e.g. by utilizing magnetic photonic crystals [ 8,9,10,11,12], cavity arrays [ 15], synthetic gauge field produced in coupled ring resonators [ 13,14,22,23], and coupled helical fibers [ 17]. A recently proposed path is to emulate the quantum spin Hall (QSH) effect [ 6,24,25,26] by introducing a photonic analog of spin-orbital interaction using bianisotropic metamaterials [ 27,28,29,30,31] and metawaveguides [ 32,33]. Another alternative is to emulate the quantum valley Hall (QVH) effect [ 34,35,36,37,38] by breaking the inversion symmetry of the system. A particularly promising design [ 39] stands out in the context of utilizing PTIs devices as applications owing to the compatibility of various topological phases on a single platform. Such a structure implementing with multiple independent binary degrees of freedoms (DOFs) is evidently capable of providing functionalities that one binary DOF could not achieve, for example, spin/valley waves sorting, polarized circulator and random cavity.

In this Letter, we experimentally demonstrate a valley sorting platform (see Fig. 1) based on spin-valley-locked edge states at the heterogeneous interface between two topologically distinct photonic crystals (PCs), namely spin photonic crystals (SPCs) and valley photonic crystals (VPCs) which emulate photonic analogue of QSH effect and quantum valley Hall (QVH) effect respectively.

The topologically protected edge waves at a heterogeneous interface between SPC and VPC have been constructed with a 4-band model in the previous studies [ 32,40,39]. The design principles, however, are briefly presented here for the sake of completeness and all dimensional parameters are rescaled according to the mid-bandgap frequency $f_0 = 6.08$GHz for practical purposes. To encompass two different DOFs in one platform, a judicious design of PC Fig1, a hexagonal lattice of metallic cylinders with symmetric air gaps at top and bottom ends sandwiched by a metallic parallel plates waveguide, is considered as a common unperturbed starting point of both SPC and VPC. This simple waveguide is referred as a photonic graphene due to the linear dispersion of transverse electric (TE) and transverse magnetic (TM) modes at $K$ and $K'$ points in photonics band structure (PBS) that resembles to the Dirac cone of electron dispersion in graphene. These Dirac point crossings can be destroyed by a broad class of perturbations that creates a complete photonic bandgap (PBG). The resulting gapped PC can, in fact, support electromagnetic wave propagation with topological properties under a subclass of perturbations that preserve specific DOFs including the synthetic spin DOF and the valley DOF at the vicinity of Dirac points.

In the case of SPC, the spin DOF is constructed from the in-phase and out-of-phase relation between degenerate TE and TM modes at $K/K'$ points. The spin-preserving perturbation which emulates the SOC in solid state physics is introduced by connecting one end of the cylinder with the confining plate to close one of the cylinder-to-plate air gaps as in Fig. 1a. This bianisotropic perturbation breaking z mirror symmetry gives rise to a photonic bandgap centered at the double-degenerate Dirac point Fig. 1d with the spectral width proportional to the overlap integral $\Delta_{SOC}$ between TE/TM modes inside the sealed cylinder-to-plate air gap. The electromagnetic modes propagating above and below the bandgap have a QSH-like topological nature with a spin-Chern number given by $2C_{\uparrow/\downarrow,v}^{SOC} = \pm 1 \times sgn(\Delta_{SOC})$, where $s = \uparrow, \downarrow$ is the spin label, $v = K, K'$ is the valley label, and the valley-independent perturbation strength $\Delta_{SOC} > 0$ ($\Delta_{SOC} < 0$) if the bottom (top) cylinder-to-plate gaps are closed with metal filling.

Similarly, the valley-preserving perturbation of the VPC is imposed by deforming cylinder of the unperturbed design into a tripod-like, $C_3$-symmetric shapes shown in Fig. 1b that breaks the in-plane inversion symmetry with respect to the principle axes of hexagonal lattice in Fig. 1c. It is essential to match the perturbation strength of TE and TM modes so that the perturbed structure becomes VPC and henceforth supports propagating electromagnetic waves below the bandgap in Fig. 1e with a spin-independent valley-Chern number $2C^P_{s,K/K'} = \pm 1 \times sgn(\Delta_P)$, where $\Delta_P$ is the perturbation strength of the inversion symmetry breaking. The width of this bandgap is proportional to $\Delta_P$ which can be controlled by rotating the tripods with respect to the principle axes and is maximized at $\theta = 0°\ (\Delta_P > 0), 60°(\Delta_P < 0)$ where the strongest perturbation happens (see supplemental material for details). A critical step toward implementing independent DOFs onto one common platform is to match the bandgap of VPCs with the bandgap of SPCs so that the edge states could reside in the same spectral range as indicated in Fig. 1d and 1e.

Building upward from the foundation laid in the previous paragraphs, a heterogeneous interface could be constructed using SPCs and VPCs on a common platform. Indeed, such an interface locks the relation of spin- and valley- DOFs at the domain wall, thus supports specific valley-spin locked TPEWs propagation. For example, the supercell simulation in Fig. 2a suggests the domain wall between SPC with $\Delta_{SOC} > 0$ and VPC with $\Delta_P > 0$ supports wave propagation with locked down DOFs, namely spin-up wave at $K'$ point and spin-down wave at $K$ point. This locking could be easily understood using symbolic notation according to the principle of *bulk-boundary correspondence*. In the case of Fig. 2a, the only two existing edge states are a spin-up wave at $K'$ point: $C^{SOC}_{\uparrow,K'} - C^P_{\uparrow,K'} = +1/2 - (-1/2)$ and a spin-down wave at $K$ point: $C^{SOC}_{\downarrow,K} - C^P_{\downarrow,K'} = (-1/2) - (+1/2)$ since the other two combinations are zero. On the other hand, the same principle gives to a spin-up wave at $K$ point and a spin-down wave at $K'$ point at the interface between SPC with $\Delta_{SOC} > 0$ and VPC with $\Delta_P < 0$ for the interface in Fig. 2b. An important insight could be deduced from these PBSs that the dispersion relation of edge states in Fig. 2a acts like a EM wave propagates in a medium with positive refractive index as the group velocity is positive (negative) at positive (negative) $k_x$ and the other way around for the edge states in Fig. 2b. Therefore, the TPEWs will be referred as positive index mode $\Psi^{(\uparrow/\downarrow)+}$ in Fig. 2a and as negative index mode $\Psi^{(\uparrow/\downarrow)-}$ in Fig. 2b from this point on. It is also worth noticing a fact that the color plots of $|E_z|$ component in the insets are spatially dissimilar to each other for both modes. Particularly the low strength spots (blue double circles in schematics) locate differently in Fig. 2a and 2b. This provides a control handle to preferentially excite one of two modes. In other words, one can launch more energy into one mode by placing a linear antenna along z-axis at the low strength spot of the other mode. This idea will be thoroughly examined in the later part of this article.

To fully utilize both modes of TPEWs and to further demonstrate the functionality of valley waves sorting, a platform consisting of $15 \times 15$ unit cells has been proposed and fabricated in Fig. 3a. The platform is partitioned into three bordering domains of PCs, which are VPC with $\Delta_P > 0$ at $\theta = 0°$ (yellow shaded area), SPCs with $\Delta_{SOC} > 0$ (green shaded area) and $\Delta_{SOC} < 0$ (blue shaded area). The domain wall between SPC/SPC ($\Delta_{SOC} > 0, \Delta_{SOC} < 0$) guarantees the launching antenna at Port A excite spin-up waves $\Psi^{\uparrow(+/-)}$, containing both positive (red arrow) and negative index (purple arrow) modes [33], toward positive x-direction. One thing to be aware at this point is that the ratio of these two modes is undetermined since the transmitter location is not specified at Port A. For the sake of clarity, this article will first focus on the positive-preferred excitation (red double circle in Fig. 3b) and move onto the other configuration later. At the common vertex of these domains, there are two other topologically distinct SPC/VPC interfaces. The one extending along the positive x-direction is situated between $\Delta_{SOC} > 0$ and $\Delta_P > 0$ and hence supports the positive index mode $\Psi^{\uparrow+}$. Similarly, the other interfaces angled 60° from x-axis hosts the negative index mode $\Psi^{\uparrow-}$ at the domain wall between PCs with $\Delta_{SOC} > 0$ and $\Delta_P < 0$. Upon encountering with this y-shaped junction, the excited TPEW is being sorted into either SPC/VPC interfaces according to its valley DOF and is then picked up by a receiving antenna placed just outside of the platform at Port B or Port C in Fig. 3a. All ports are connected to a vector network analyzer (Agilent E5071C) to measure the *S*-parameters of valleyed-sorted TPEWs through A-B or A-C channel. The receiver at both Port B and C is mounted on a translation stage (BiSlide, VelMex Inc.) programed to move the antenna to the next location after completion of single VNA measurement so that one could acquire a series of transmission spectrum corresponding to different locations along the outer edge of the platform (gray dashed arrows). To better observe the energy routing of the platform, the measured $|S21|^2$ of spectrum is double-summed over every location along the scanned path and frequencies points in the spectral range of the photonic bandgap. The resulting transmitted energy through A-B channel (blue stars) and A-C channel (green circles) is plotted in Fig. 3c and further verifies the excited waves contain more $\boldsymbol{\Psi}^{\uparrow+}$ content using the positive-preferred excitation configuration of Port A. Noting that the purity of excited modes could be easily improved by using an array of phased dipole antennas, but this optimizing process is beyond the scope of this paper.

Another novel property of this valley-spin locked platform is the capability to binary-switch the topology of VPCs from $\Delta_P > 0$ to $\Delta_P < 0$ by rotating the tripods from $\theta = 0°$ to $\theta = 60°$. In doing so, the supported modes of two channels swapped accordingly due to the distinct topology of VPC domain. In Fig. 3c, for example, $\boldsymbol{\Psi}^{\uparrow+}$ is previously directed to Port B at $\theta = 0°$ and is now routed to Port C due to the SPC/VPC segment of A-B channel is situated between $\Delta_{SOC} > 0$ and $\Delta_P < 0$ at $\theta = 60°$ and thus suppresses the propagation of the positive index TPEWs (see supplemental information). Likewise, this switching of topology affects the SPC/VPS domain wall of A-C channel as well and eventually two channels interchange the supporting TPEWs modes. This transition of topology is monitored by gradually rotating tripods from $\theta = 0°$ to $\theta = 60°$ with $10°$ per step. The measured result clearly shows the transmitted signal decreases at Port B and increases at Port C throughout the process. The numerical simulation of this transition of topology (blue and green dashed curves) agrees well to the experimental observation and confirms the signal change is in fact due to energy routing in the insets at $= 0°$ and $\theta = 60°$. Combining above demonstrations, the platform is shown to not only have the capability of valley waves sorting, but also be able to dispatch the TPEWs of specific DOFs to the desired port.

With clear display of platform characteristics with positive-preferred excitation, we now shift the focus to the controlling of mode excitation mentioned in the earlier context by studying the case of negative-preferred excitation. Using a launcher at purple double circle in Fig. 3b, one can more efficiently excite the negative index mode. In Fig. 3d, one would expect the launched waves with predominately $\boldsymbol{\Psi}^{\uparrow-}$ mode will be sorted and dispatched in a simply opposite way compared to the wave with majorly $\boldsymbol{\Psi}^{\uparrow+}$ mode. However, an obvious discrepancy happens at A-B channel with lower transmitted energy roughly by a factor of 2. This seemingly abnormal asymmetry could be readily explained by the outcoupling behavior of valley-polarized waves at various terminations in the recent reports [40,41]. In short, the valley DOF plays an important role in suppressing the back-scattering from the zigzag termination (A-C channel) of crystals and subsequently granting efficient outcoupling into the ambient environment. In contrast, the arm-chair termination (A-B channel) could reflect TPEWs and deteriorate outcoupling efficiency since the valley DOF is not preserved at the edge of the platform. It is important to also notice a fact that although both $\boldsymbol{\Psi}^{\uparrow+}$ and $\boldsymbol{\Psi}^{\uparrow+}$ suffer from the inefficient outcoupling in A-B channel, the arm-chair termination is more deleterious to the negative index mode than to the positive index mode due to larger impedance mismatch (see supplemental information for details).

In conclusion, we have experimentally demonstrated a valley-spin locked platform based on PTIs in radio frequency. Implementing SPCs and VPCs into a single structure enables numerous applications, such as valley waves sorting and polarized waves dispatching, that is impossible to construct from PCs with only one binary DOF. The predicted topological transition of VPCs is observed and later utilized to substantiate valley-polarized energy routing. The crucial role of valley conservation on this platform is directly examined by studying the efficiency of TPEWs outcoupling at crystal termination. With the functionalities of this design firmly established in this study, this design could pave the way toward real world applications based on PTIs. Aside from the down-to-earth interest, this valley-spin platform naturally serves as an ideal test bed for emulating phenomenon in the context between spintronics and valleytronics.

**Figures**

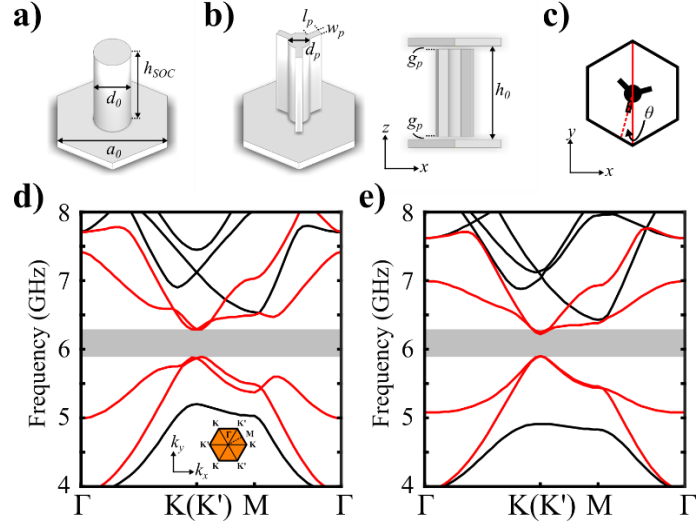

**Figure 1.** Designs of photonic crystal with non-trivial topology. **a)** Spin photonic crystal (SPC) emulating Quantum Spin Hall effect with broken mirror symmetry in z-axis. **b)** Valley photonic crystal (VPC) emulating Quantum Valley Hall effect with broken in-plane inversion symmetry with respect to primitive axes of hexagonal lattice. **c)** Schematic unit cell of VPC illustrating rotation angle $\theta$ between the tip of tripod (red dashed line) and negative y-direction (red solid line). The symmetry-breaking perturbation is maximized at $\theta = 0°$ ($\Delta_P > 0$), $60°$ ($\Delta_P < 0$). **d)** and **e)** Calculated photonic band structures of **d)** SPC and **e)** VPC at $\theta = 0°$. Red curves: topologically non-trivial modes; black curves: dispersion of bulk modes; gray shaded area shows a complete photonic bandgap centered at $f_0 = 6.08$ GHz with 7% bandwidth. The bandgap of VPC is matched to the bandgap of SPC by tweaking symmetry-breaking perturbation with the dimensions of tripod. Geometric parameters: $a_0 = h_0 = 36.81mm, h_{SOC} = 31.29mm, d_0 = 12.70mm, l_p = 4.27mm, w_p = 2.21mm, d_p = 7.36mm, g_p = 1.10mm$.

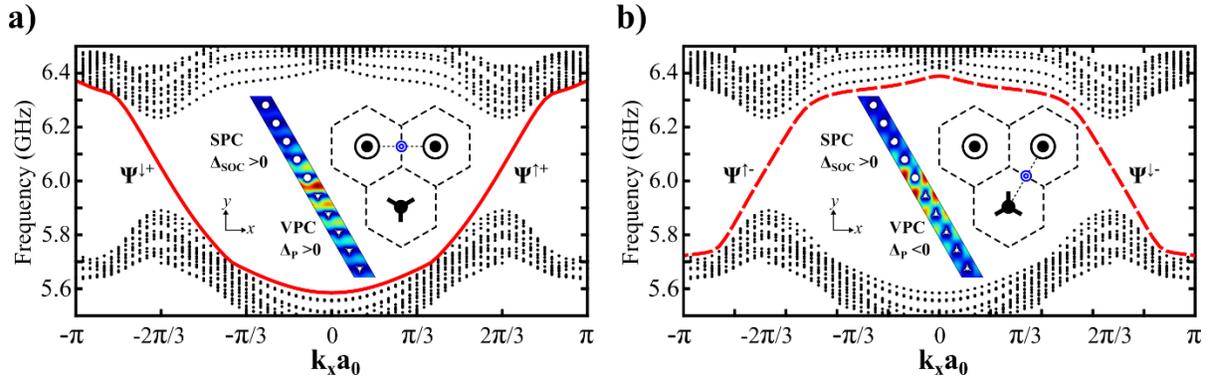

**Figure 2.** Topologically protected edge waves (TPEWs) with Valley-Spin locking propagation. Photonic band diagram of a heterogeneous interface between **a)** SPC with $\Delta_{SOC} > 0$ and VPC with $\Delta_P > 0$ and **b)** SPC with $\Delta_{SOC} > 0$ and VPC with $\Delta_P < 0$. The band structure is calculated with a supercell with periodicity $a_0$ along x-direction and 15 cells on each side of the domain wall. Red solid curves: positive index TPEW; red dashed curves: negative index TPEW; black dots: bulk modes; edge modes are denoted with positive/negative (+/−) indices and up/down (↑/↓) spins. Inset: $|E_Z|$ component for two kinds of SPC-VPC interfaces in supercell simulation (only 10 cells are included in figure for simplicity); blue double circles: locations of low $|E_Z|$ value at the vicinity of domain wall for preferential mode excitation.

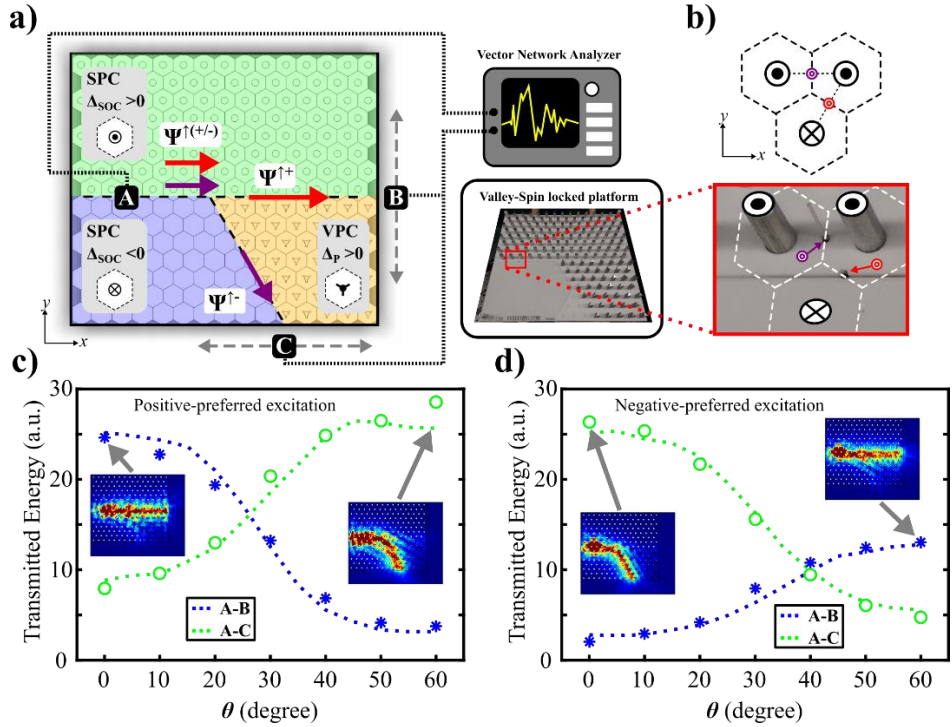

**Figure 3.** Demonstration of valley wave sorting on a Valley-Spin locked platform. **a)** Schematic of radio frequency test station. The launching antenna (Port A) sticks into the structure through a drilled hole located $2.5a_0$ from leftmost edge on the bottom plate. The receiving antenna is mounted on a translation stage to scan along the outer edges of the platform at Port B or Port C. The transmitted signal through channel A-B or A-C is measured by a vector network analyzer. Inset: the fabricated platform with top plate removed to reveal the layout of PCs. **b)** Schematic and picture of Port A indicating preferential mode excitation using different launching locations. Red double circle: positive-preferred excitation; purple double circle: negative-preferred excitation. **c)** and **d)** Transmission measurements of the Valley-Spin locked platform with **c)** positive-preferred excitation and **d)** negative-preferred excitation. Blue star (green circle) shows the summation of experimentally measured $|S21|^2$ through A-B (A-C) channel. Numerical simulation of the transmission through A-B (blue dashed curve) and A-C (green dashed curve) channels agrees well to experimental observations. The tripods are rotated gradually from $\theta = 0°$ to $\theta = 60°$ (from $\Delta_P > 0$ to $\Delta_P < 0$) with $10°$ per step. Insets: energy density plots of full-size simulation at $\theta = 0°$ and $\theta = 60°$ showing the energy is in fact routed according to the topology swapping of two SPC-VPC interfaces.

# Supplemental Material: Valley Waves Sorting and Routing on a Valley-Spin Locked Platform Based on Photonic Topological Insulators

Kueifu Lai[1,3], Yang Yu[1], Yuchen Han[2], Fei Gao[4], Baile Zhang[4], and Gennady Shvets[1*]

[1]School of Applied and Engineering Physics, Cornell University, Ithaca NY 14853, USA.

[2]Department of Physics, Cornell University, Ithaca NY 14853, USA.

[3]Department of Physics, University of Texas at Austin, Austin, TX 78712, USA.

[4]Division of Physics and Applied Physics, School of Physical and Mathematical Sciences, Nanyang Technological University, Singapore 637371, Singapore.

* Author to whom correspondence should be addressed; E-mail: gs656@cornell.edu (G. Shvets)

**Sample fabrication and assembly**

The aluminum tripod of VPC is fabricated using wire electrical discharge machining (wire EDM). The structural foam (ROHACELL 51HF) is used on both ends to support the tripod in the center of parallel plates waveguide. Another piece of structural foam is laser-cut into a mask to orient the direction of tripods.

**Data processing of VNA measurements**

A LabView program is developed to automatically retrieve data from VNA and move the receiving antenna a step further for the next data acquisition. A complete set of measurements contains 257 transmission spectra taken at designated locations along the edge of platform. The measured $|S21|^2$ of transmission spectrum is firstly summed over 257 locations to acquire the total transmitted energy of one edge. The resulting spectra is then summed at spectral positions inside the photonic bandgap, *i.e.* for frequency points in the range $5.9 \text{ GHz} < f < 6.3 \text{GHz}$ to better represent the energy routing of TPEWs.

**Numerical simulation configuration**

The photonic band structures and the source-driven simulations are calculated by the first-principles simulation package *COMSOL Multiphysics* using radio frequency module. To better reproduce experimental conditions in the driven simulation, a point dipole along z-direction is placed at the $z = 0.5h_0$ serving the excitation source at Port A and a domain of ambient air is considered along the edges of the structure for modelling TPEWs outcoupling. The said model is simulated at 5 spectral positions spanning evenly from $0.98f_0$ to $1.02f_0$ and the transmitted energy is calculated by integrating $|E_z|^2$ at the exiting surface of Port B and Port C. The resulting value are summed over 5 frequencies inside PBG as a single measure to match the experimental results.

## Controlling photonic bandgap of VPC unit cell

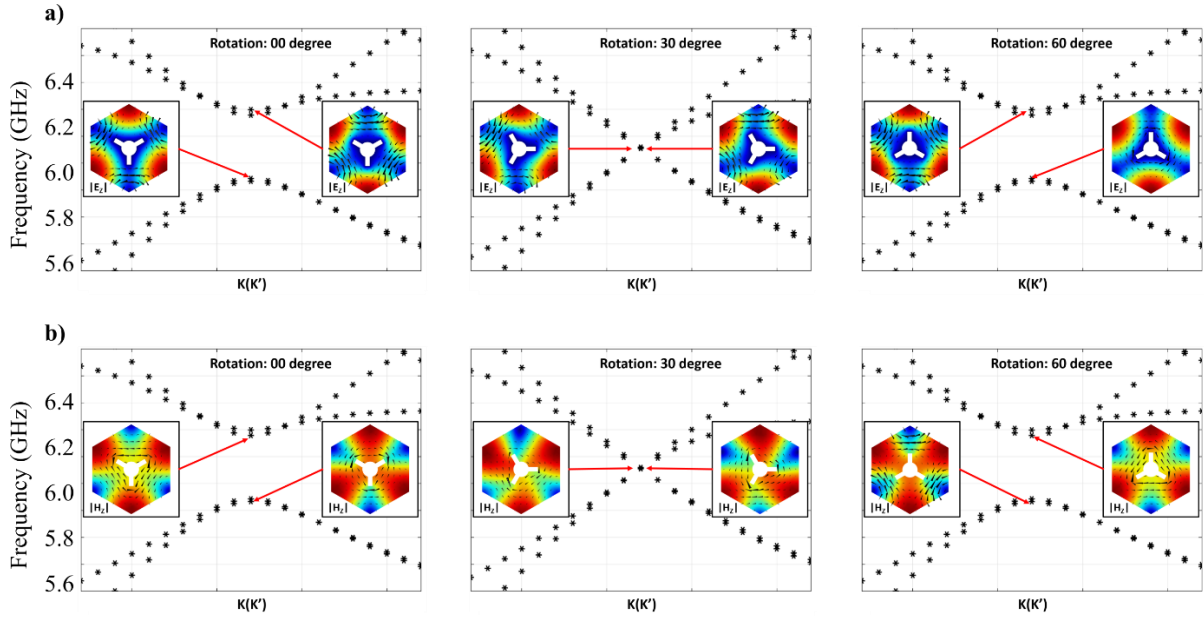

**Figure S1.** Photonic band structures of VPC unit cells with various perturbation strength. Field profiles of the relevant bands at $K/K'$ points for **a)** $|E_Z|$ component of TE modes and **b)** $|H_Z|$ component of TM modes at $\theta = 0°, 30°, 60°$. Color plot: strength of field component; black arrow: Poynting flux.

The strength of the symmetry-breaking perturbation can be controlled by the rotation angle $\theta$. The series of figures clearly shows the bandgap is maximized at $\theta = 0°, 60°$ and closed at $\theta = 30°$ where the in-plane inversion symmetry is not broken. It is worth noticing that the bands flip after the bandgap closed and opened giving due to the change of topology.

## Valley-polarized wave routing on the valley-spin locked platform

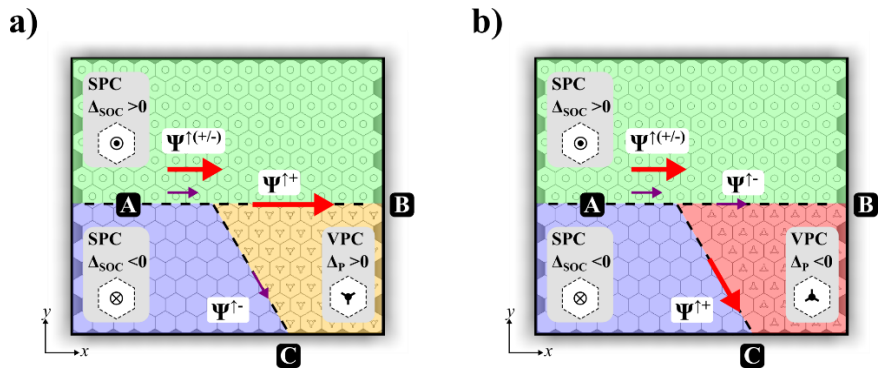

**Figure S2.** Energy routing of the platform with positive-preferred excitation. Schematic of valley-polarized waves routing **a)** with VPC at $\theta = 0°$ ($\Delta_P > 0$). **b)** with VPC at $\theta = 60°$ ($\Delta_P < 0$). Green: SPC with $\Delta_{SOC} > 0$, blue: SPC with $\Delta_{SOC} < 0$, yellow: VPC with $\Delta_P > 0$, red: VPC with $\Delta_P < 0$.

Schematics of the energy flow on valley-spin locked platform. The supported modes at two SPC/VPC interfaces swaps due to the topology of the VPC domain change from $\Delta_P > 0$ to $\Delta_P < 0$.

## Valley protected TPEWs outcoupling

The TPEW outcoupling protected by valley DOFs is further demonstrated on the other valley-spin locked platform with zigzag terminations at both Port B and Port C as shown in the Fig. S3a.

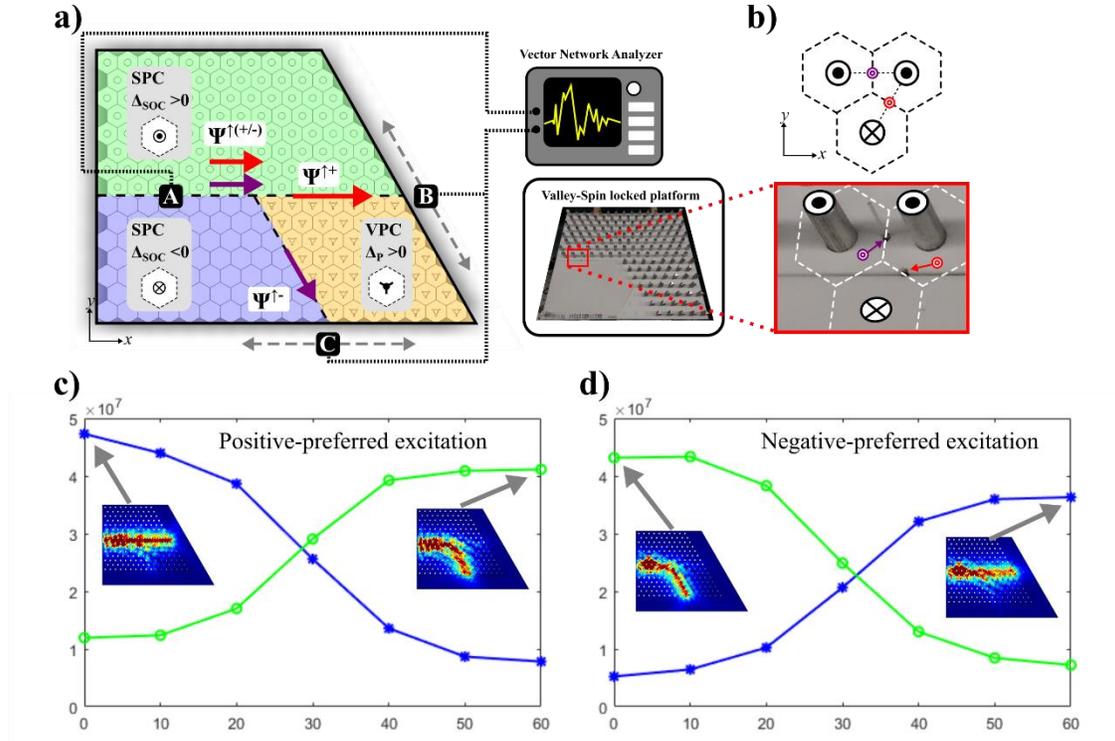

**Figure S3.** Numerical simulation of valley-spin locked platform with two zigzag terminations. **a)** Alternative layout of valley-spin locked platform with efficient outcoupling. **c)** and **d)** Measured and simulated transmitted energy using **c)** positive-preferred excitation **d)** negative-preferred excitation.